\documentclass[fleqn,usenatbib]{mnras}

\usepackage[T1]{fontenc}

\DeclareRobustCommand{\VAN}[3]{#2}
\let\VANthebibliography\thebibliography
\def\thebibliography{\DeclareRobustCommand{\VAN}[3]{##3}\VANthebibliography}

\usepackage{graphicx}	
\usepackage{amsmath}	
\usepackage{amssymb}	
\usepackage{multirow}
\usepackage{caption}
\usepackage{multicol}        
\usepackage{longtable}
\usepackage{xr}
\usepackage{newtxtext,newtxmath}

\title[FRB 20121102A]{FRB 20121102A: images of the bursts and the varying radio counterpart}

\author[L. Rhodes et al.]{L. Rhodes$^{1}$\thanks{E-mail: lauren.rhodes@physics.ox.ac.uk}
M. Caleb$^{2, 3}$
B. W. Stappers$^{4}$
A. Andersson$^{1}$
M.C. Bezuidenhout$^{4}$
L. N. Driessen $^{2}$
\newauthor
I. Heywood $^{1,5,6}$
\\
$^{1}$ Astrophysics, Department of Physics, University of Oxford, Denys Wilkinson Building, Keble Road, Oxford, OX1 3RH, UK\\
$^{2}$ Sydney Institute for Astronomy, School of Physics A28, University of Sydney, NSW 2006, Australia\\
$^{3}$ ASTRO3D: ARC Centre of Excellence for All-sky Astrophysics in 3D, ACT 2601, Australia\\
$^{4}$ Jodrell Bank Centre for Astrophysics, Department of Physics and Astronomy, The University of Manchester, Manchester, M13 9PL, UK\\
$^{5}$ Centre for Radio Astronomy Techniques and Technologies, Department of Physics and Electronics, Rhodes University,
PO Box 94, Makhanda 6140, South Africa\\
$^{6}$ South African Radio Astronomy Observatory, 2 Fir Street, Black River Park, Observatory, Cape Town, 7925, South Africa\\
}

\date{Accepted XXX. Received YYY; in original form ZZZ}

\pubyear{2015}

\begin{document}
\label{firstpage}
\pagerange{\pageref{firstpage}--\pageref{lastpage}}
\maketitle

\begin{abstract}
As more Fast Radio Bursts (FRBs) are being localised, we are learning that some fraction have persistent radio sources (PRSs). Such a discovery motivates an improvement in our understanding of the nature of those counterparts, the relation to the bursts themselves and why only some FRBs have PRSs. We report on observations made of FRB 20121102A with the MeerKAT radio telescope. Across five epochs, we detect the PRS associated with FRB 20121102A. Our observations are split into a cluster of four epochs (MJD 58732 - 58764) and a separate single epoch about 1000\,days later. The measured flux density is constant across the first four observations but then decays by more than one-third in the final observation. Our observations on MJD 58736 coincided with the detections of 11 bursts from FRB 20121102A by the MeerTRAP backend, seven of which we detected in the image plane. We discuss the importance of image plane detections when considering the commensal transient searches being performed with MeerKAT and other radio facilities. We find that MeerKAT is so sensitive that within a two-second image, we can detect any FRB with a flux density above 2.4\,mJy at 1.3 GHz and so could localise every FRB that has been detected by CHIME to date.
\end{abstract}




\section{Introduction}

Fast Radio Bursts (FRBs) are flashes of coherent radio emission lasting around a few to hundreds of milliseconds \citep[e.g.][]{2007Sci...318..777L, 2012MNRAS.425L..71K, 2016PASA...33...45P}. The dispersion measure (DM), the result of electrons along the line of sight delaying the arrival of lower frequency photons, can be used as a proxy for distance and has implied that FRBs must originate from cosmological distances \citep{2013Sci...341...53T}.

There are many theories as to the origin of FRBs \citep[][]{2020Natur.587...45Z, 2022arXiv221203972Z} but the recent discovery of pulsed radio emission from Soft Gamma-ray Repeater (SGR) 1935+2154 by the STARE2 and CHIME telescopes suggests that magnetars like SGR 1935+2154 may produce some of the population of FRBs \citep{2020Natur.587...59B, 2020Natur.587...54T}. The luminosity of the bursts from SGR 1935+2154 overlap with the low end of the luminosity distribution of cosmological FRBs \citep{2020Natur.587...54T, 2020Natur.587...59B}. Surveys designed to monitor the sky to find FRBs have discovered that at least a subset of the population repeat \citep{2019ApJ...885L..24C, 2020ApJ...891L...6F}. Such a discovery has provided the community with another clue as to the origin of FRBs: at least part of the population cannot originate from cataclysmic events. FRB 20121102A was the first known repeating FRB \citep{2016Natur.531..202S, 2016ApJ...833..177S}. \citet{2020MNRAS.495.3551R} and \citet{2021MNRAS.500..448C} first showed that this source has active periods occurring every $\sim$160\,days confirming that some FRBs are produced by objects in binaries or are processing \citep{2020Natur.582..351C, 2020ApJ...895L..30L, 2020ApJ...893L..39L, 2020ApJ...892L..15Z}.

FRB 20121102A was also the first FRB to be localised. \citet{2017Natur.541...58C} were able to constrain the position of FRB 20121102A to sub-arcsecond levels using the Karl G. Jansky Very Large Array (VLA). The position was coincident with a dwarf galaxy at $z = 0.19$ \citep[luminosity distance of 926\,Mpc for $H_0 = 70$\,km/s/Mpc, $\Omega_{M} = 0.3$][]{2017ApJ...834L...7T}. The position of the FRB is coincident with a persistent radio source (PRS). High angular resolution radio observations show the PRS to be unresolved on all scales and constrain the source size to be $\lesssim$ 0.7pc \citep{2017ApJ...834L...8M}.

The PRS associated with FRB 20121102A has now been detected between 100\,MHz and 30\,GHz \citep{2017Natur.541...58C, 2020arXiv201014334R} and its spectral shape is best described by a broken power law with a break around 8\,GHz. Long-term monitoring has shown variability on day-timescales at a level of 10\% \citep{2017Natur.541...58C}, which has since been attributed to scintillation \citep{2017ApJ...842...34W}. 


The leading theory as to the origin of the PRS is a pulsar wind nebula (PWN) containing a highly magnetised pulsar i.e. a magnetar \citep[][]{2017ApJ...834L...8M, 2017ApJ...841...14M, 2019ApJ...885..149Y}. A PWN would be produced due to the confinement of a young magnetar's wind within the environment of supernova remnant (SNR) in which the magnetar is encased \citep[see][for a review]{2006ARA&A..44...17G}. As the supernova expands and evacuates the local interstellar medium, the magnetar produced in the supernova powers a nebula inside the evacuated region. The spin down energy of the magnetar powers the expansion of the nebula out to the SNR and the highly energetic particles in the wind are responsible for the persistent radio emission. Successful attempts to reproduce observations using a pulsar wind model have been made, demonstrating that is a viable origin for the observed persistent radio emission \citep{2019ApJ...885..149Y, 2017ApJ...841...14M}. Studies of the persistent source's spectrum show that the radio emission is still very young. The spectral break at 8\,GHz has led to the calculation of an approximate PRS age of a few hundred years \citep{2020arXiv201014334R}. A young system would also explain why the PWN is still so luminous. Despite the extensive study of the PRS associated with FRB 20121102A and others; it is still not known if these systems are separate and distinct from the population of FRBs without PRSs \citep{2022ApJ...927...55L}.


In this paper, we present observations of the persistent radio source associated with FRB 20121102A with the MeerKAT telescope. In Section \ref{sec:observations} and \ref{sec:results}, we describe the observations that were made and the results of these observations including image plane detections of seven bursts previously reported in \citet{2020MNRAS.496.4565C}, respectively. In Section \ref{sec:discussion} we discuss how our results fit with previous observations and the implications of detecting FRBs with MeerKAT for future commensal searches.

\section{Observations}
\label{sec:observations}

Observations with MeerKAT were obtained through a Director's discretionary time proposal (DDT-20190905-MC-01, PI: M. Caleb). The MeerKAT radio telescope is a 64-dish radio interferometer based in the Karoo desert in South Africa \citep{2016mks..confE...1J}. A total of five observations were made on 2019 September 6\textsuperscript{th}, 10\textsuperscript{th}; October 6\textsuperscript{th}, 8\textsuperscript{th}, and 2022 September 26\textsuperscript{th} (MJDs 58732, 58736, 58762, 58764 and 59848, respectively). Data were taken at a central frequency of 1.28\,GHz with a bandwidth of 856\, MHz, which was split into 4096 channels for calibration. The observations were made with 8-second integrations with the exception of the 10\textsuperscript{th} September which used 2-second integrations. J0408-6565 and J0534+1927 were used as the primary and secondary calibrators for each observation, respectively. Calibration and flagging were performed in \textsc{casa} \citep[Common Astronomy Software Applications][]{casa}. Before calibration, flagging was performed on the calibrator scans. Bandpass and phase-only gain calibration was performed on the primary. Complex gain calibration was performed on both calibrators after which a flux scale was applied, using the known model of the primary, to the secondary. The corrections were then applied to the target field.  The calibrated target fields were flagged to remove RFI with \textsc{tricolour} \citep{2022ASPC..532..541H} and imaged using \textsc{wsclean} \citep{2014MNRAS.444..606O}. Imaging used a Briggs weighting of -0.3. The declination of the FRB 20121102A is far north with respect to the MeerKAT observing range and so the synthesised beam in each image is particularly elongated. The field was imaged in two separate spectral bands to calculate a two-point spectral index of the PRS (the source was too faint to detect the PRS with a higher spectral resolution). The flux density and spectral index associated with the PRS in each observation are reported in Table \ref{tab:fluxes}. Any reported flux densities were measured using the \textsc{casa} task \textit{imfit}.


\begin{table}
\centering
 \caption{A table presenting the PRS flux densities and rms noise values in the five FRB 20121102A MeerKAT observations. The uncertainty on the flux is a combination of the statistical error and a 10\% calibration error added in quadrature. The large uncertainties on the spectral index measurements are a combination of the relatively small bandwidth over which to make the measurement as well as the low flux density of the PRS.}
 \label{tab:fluxes}
 \begin{tabular}{cccc}
  \hline
  Date (mjd) & Observation duration (hrs) & S\textsubscript{$\nu$} ($\mu$Jy/beam) & $\alpha$ (S\textsubscript{$\nu$}$\propto\nu^{\alpha}$) \\
  \hline
  58732.16 & 3.0 & 260$\pm$26 & -0.6$\pm$1.0\\
  58736.13 & 3.0 & 269$\pm$27 & -0.5$\pm$1.0\\
  58762.10 & 2.5 & 287$\pm$27 & -0.1$\pm$0.4\\
  58764.10 & 3.0 & 270$\pm$28 & -0.4$\pm$1.0\\
  59848.07 & 3.0 & 189$\pm$18 & -0.8$\pm$0.6\\
  \hline
 \end{tabular}
\end{table}

\section{Results}
\label{sec:results} 

\subsection{Persistent Radio Source}
\begin{figure*}
    \centering
    \includegraphics[width=\textwidth]{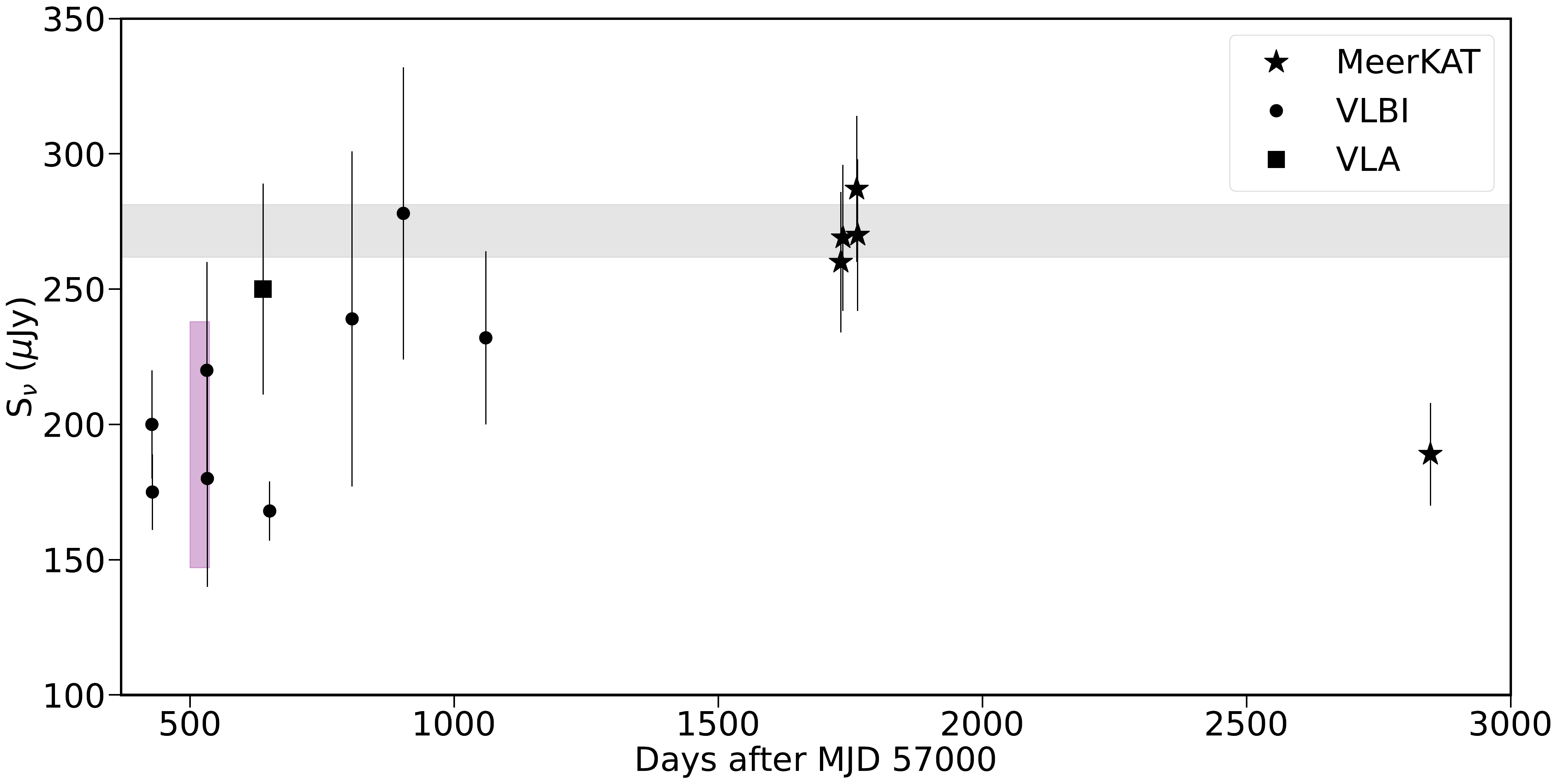}
    \caption{The radio light curve of the PRS associated with FRB 20121102A at L-band \citep[Table \ref{tab:fluxes}, ][]{2017ApJ...834L...8M, 2022MNRAS.511.6033P}. The grey horizontal region denotes the average ($\pm$1 standard deviation) flux density as measured by MeerKAT between MJD 58732 and 58764. The purple highlighted region denotes the period where significant 3\,GHz variability was observed, the vertical extent indicates the maximum and minimum 3\,GHz flux density measured \citep{2017Natur.541...58C}.}
    \label{fig:lightcurve}
\end{figure*}

The final images showed a point source at the phase centre of the field, its position consistent with those reported for previous detections of the PRS associated with FRB 20121102A \citep{2017ApJ...834L...8M, 2023MNRAS.518.3462A}. The PRS was clearly detected, with a signal-to-noise (S/N) $>10\sigma$, in each observation. There is no significant variability in the measured flux density of the PRS during the first four observations (2019). However, between the 2019 observing campaign and the 2022 observation, there is a significant drop in the PRS's flux density. The flux density over the three-year period decreases by about one-third. The flux densities are given in Table \ref{tab:fluxes} and Figure \ref{fig:lightcurve}.

\begin{figure}
    \centering
    \includegraphics[width = \columnwidth]{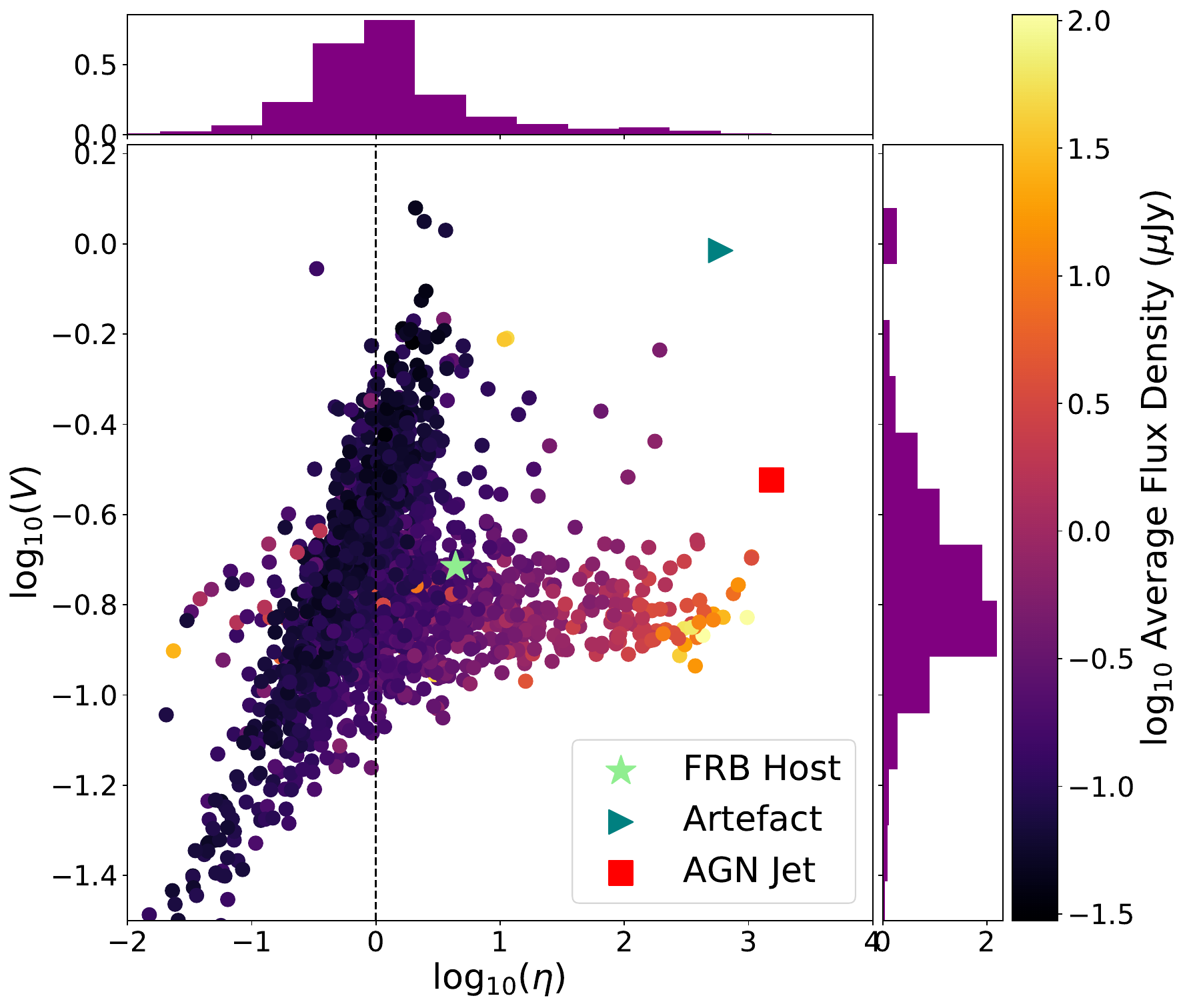}
    \caption{A plot of the amplitude of the variability (V) as a function of the statistical significance of the variability ($\eta$) for every source within the FRB 20121102A MeerKAT field. On the top and right-hand side of the plot are two histograms of the two parameters. The colour scale of each data point denotes the average flux density of each source. Three sources are highlighted: the green star denotes the FRB host i.e. the PRS, the triangle is an imaging artefact and the square is a resolved jet whose variability is correlated with the synthesised beam shape.}
    \label{fig:etav}
\end{figure}

To ensure that the flux density reduction of the PRS is real and not a calibration error, we used the TRAnsients Pipeline \citep[TraP,][]{2015A&C....11...25S} to search for any systematic changes in flux density across the whole field over the five observations. For each source, we calculated $\eta$ (Equation \ref{eq:eta}), which is analogous to a $\chi^2$ statistic by comparing a given source to a model with a constant flux density, and $V$ (Equation \ref{eq:V}) which is a measure of the spread of flux densities for a given source. $\eta$ and $V$ are defined as

\begin{equation}\eta_v \equiv \chi_{N-1}^2=\frac{1}{N-1} \sum_{i=1}^N \frac{\left(F_{i, v}-{\overline{F_v}}^*\right)^2}{\sigma_i^2}\label{eq:eta}\end{equation}

\begin{equation}V_v \equiv \frac{s_v}{\overline{F_v}}=\frac{1}{\overline{F_v}} \sqrt{\frac{N}{N-1}\left(\overline{F_v^2}-{\overline{F_v}}^2\right)}\label{eq:V}\end{equation}

where $F_{i, v}$ is the flux density of a source at frequency $\nu$ in observation $i$ , $\overline{F_v}^*$ is the weighted mean flux density of a source across all observations, $\overline{F_v}$ is the non-weighted mean, $\sigma_i$ is the uncertainty associated with the measured flux density and $N$ is the number of observations. The results of the TraP are shown in Figure \ref{fig:etav}. Each point corresponds to the values of $\eta$ and $V$ for a given source within the FRB field. The position of the PRS in Figure \ref{fig:etav} is denoted with a green star. Also labelled with a triangle and square are an artefact and resolved radio galaxy, respectively. Most of the sources have $\eta \leq 1$ which corresponds to non-variable objects or objects that are only detected (at low significance) in a single epoch. The horizontal structure around $\log_{10}V \lesssim -0.8$ corresponds to variability at the level of the calibration uncertainty. Sources with high $V$ and low $\eta$ are usually imaging artefacts. High flux density sources can have large $\eta$, as $\eta$ is approximately correlated with signal-to-noise squared (see Equation \ref{eq:eta}). Any small deviations in flux density, be they astrophysical or instrumental, can inflate $\eta$ for bright sources for whom the statistical uncertainty is small. The distribution of the points in the $\eta$-V plane shows no significant evidence of strong systematics introduced by the calibration process \citep{2022MNRAS.513.3482A,2022MNRAS.512.5037D}.

The histograms at the top and right-hand sides of Figure \ref{fig:etav} are approximately Gaussian in shape. Any global reduction in flux density in the final epoch would appear as strong deviations from the Gaussian-shaped histograms. Given the Gaussian shape, we can conclude that there is no evidence of a global flux density reduction in the final epoch with respect to the other four which makes the flux density reduction observed credible. 

The high S/N of the PRS detections allowed for two-point spectral index measurements. Across the five observations, we calculated a weighted mean spectral index of $\alpha = -0.4\pm0.5$, (where $S_{\nu}\propto \nu^{\alpha}$). The spectral index did not change significantly when the flux density of the PRS decreased.

\subsection{FRB detection in the image plane}\label{sub:result_FRB}

\citet{2020MNRAS.496.4565C} and \citet{2021MNRAS.505.3041P} report on the detection and analysis of bursts during this observing campaign (on 2019 September 10\textsuperscript{th}) via the MeerTRAP backend\footnote{https://www.meertrap.org} which searches for pulsars and fast transients whilst piggybacking on the large survey programmes at MeerKAT \citep{2016mks..confE..10S}. A total of 11 bursts were detected over a period of 2.5\,hours. \citet{2023MNRAS.518.3462A} presents a methodology for detecting bursts from FRB 20121102A in the image-plane. Like \citet{2023MNRAS.518.3462A}, we also imaged each two-second integration where a burst was detected and found that three bursts were detected when imaging the whole band at once. The flux densities are given in the final column of Table \ref{tab:bursts}. The emission detected from FRBs is heavily frequency-dependent. Therefore, we split the band into two halves to search for bright emission in either subband. We found that the noise was higher in the lower half of the band $\sim0.6$\,mJy/beam at 1.06\,GHz compared to $\sim0.3$\,mJy/beam in at 1.5\,GHz. The increased noise at lower frequencies is a result of more RFI in that part of the band and so a higher fraction of the data is flagged out. We detect a further four bursts in one half of the band. The increased noise makes detections at lower frequencies more difficult, only one of the four additional detections is on the lower half of the band. Where we don't detect a counterpart across the whole band but do in one subband, we provide peak flux densities and 3$\sigma$ upper limits where appropriate. In total, we detected seven of the 11 bursts reported in \citet{2020MNRAS.496.4565C}. 

\begin{table*}
\centering
\begin{tabular}{ccccc} 
    \hline
    Burst &  Frequency of subband (GHz) & S\textsubscript{$\nu$} (mJy/beam) &  Central frequency (GHz) & S\textsubscript{$\nu$} (mJy/beam)  \\
    \hline
        \multirow{2}{2em}{3} & 1.1 & $<1.8$ & \multirow{2}{2em}{1.3} & \multirow{2}{4em}{$1.4\pm0.2$} \\
      & 1.5&$1.32\pm0.2$ & & \\
          \hline
    \multirow{2}{2em}{5}& 1.1 & $<1.8$ & \multirow{2}{2em}{1.3} & \multirow{2}{3em}{$<0.9$}\\
      & 1.5&$2.4\pm0.1$ & & \\
          \hline
     \multirow{2}{2em}{7}& 1.1 & $2.3\pm0.2$ & \multirow{2}{2em}{1.3} & \multirow{2}{3em}{$<0.9$}\\
      & 1.5 & $<1.2$ & & \\
          \hline
     \multirow{2}{2em}{8}& 1.1 & $<1.8$ & \multirow{2}{2em}{1.3}& \multirow{2}{4em}{$1.6\pm0.2$} \\
      & 1.5&$1.8\pm0.1$ & & \\
          \hline
     \multirow{2}{2em}{9}& 1.1 & $<1.8$ & \multirow{2}{2em}{1.3} & \multirow{2}{3em}{$<0.9$}\\
      & 1.5&$1.9\pm0.2$ & & \\
          \hline
     \multirow{2}{2em}{10} & 1.1 & $<1.8$ & \multirow{2}{2em}{1.3}& \multirow{2}{4em}{$1.5\pm0.2$} \\
     & 1.5&$1.46\pm0.07$ & & \\
         \hline
     \multirow{2}{2em}{11} & 1.1 & $<1.8$ & \multirow{2}{2em}{1.3} & \multirow{2}{3em}{$<0.9$}\\
      & 1.5&$1.03\pm0.07$ & & \\
\hline
\end{tabular}
    \caption{A table containing all the peak flux densities of the bursts detected in either subband along with 3$\sigma$ upper limits in the case where the burst is not detected in that half of the band. We also show the flux densities for the bursts that are detected across the whole band. Whilst we detect seven bursts in either subband, only three are detected when the full band is imaged.}
    \label{tab:bursts}
\end{table*}

\section{Discussion}
\label{sec:discussion}
\subsection{The persistent radio counterpart}

The average MeerKAT flux density of the PRS: $\sim260$\,$\mu$Jy/beam, which corresponds to a radio spectral luminosity at 1.28\,GHz of 2.6$\times 10^{29}$\,erg\,s\textsuperscript{-1}\,Hz\textsuperscript{-1} (or $3.3\times10^{38}$\,erg\,s\textsuperscript{-1}) at a luminosity distance of 926\,Mpc \citep{2017ApJ...834L...7T}. 
Such a high radio luminosity is appropriate for what has been theorised for a high magnetic field (magnetar) PWN ($\sim 10^{38}$\,erg/s), thus supporting the theory that this persistent source may instead be produced by a young, highly magnetized pulsar, whose stronger magnetic field could produce more luminous radio emission \citep{2019ApJ...885..149Y, 2018MNRAS.477L..66M, 2011MNRAS.410..381B, 2008A&A...487.1033D}. 

At the time of writing, there are four other localised FRBs with associated persistent radio emission: FRBs 20201124A, 20190714A, 20190520B and 20210405I, however not all are confirmed PRSs \citep{2022MNRAS.515.1365C, 2022MNRAS.513..982R, 2022Natur.606..873N, 2023arXiv230209787D}. FRB 20210405I is most likely associated with the host galaxy as a whole and not only the FRB \citep{2023arXiv230209787D}. The source associated with FRB 20201124A is most likely to be a result of star formation because observations with the EVN (European VLBI Network) show some resolved structure \citep{2021ATel14603....1M}. FRBs 20190714A and 20190520B both have confirmed PRSs with very similar luminosities to FRB 20121102A: $1\times10^{29}$ and $3\times10^{29}$\,erg/s/Hz, respectively \citep{2022MNRAS.515.1365C, 2022Natur.606..873N}. The host galaxy of FRB 20190520B has a very low star formation rate ruling out star formation as the likely origin of the radio emission. 

A PWN would produce a spectral index around $-0.5 < \alpha < 0$ \citep[$S \propto \nu^{\alpha}$][]{1984ApJ...278..630R, 2018MNRAS.477L..66M,2017ApJ...841...14M}. We measure a weighted average spectral index of $-0.4\pm0.5$ which is in agreement with that expected for such a nebula.

Our observations of the PRS associated with FRB 20121102A show a reduction in flux density at 1.3\,GHz of about 30\% over three years. This motivates a search for L-band observations of the PRS in the literature. Figure \ref{fig:lightcurve} shows the L-band flux density as a function of time between January 2016 (MJD 57400) and November 2022 (MJD 59850). The black circles denote observations made with VLBI facilities \citep{2017ApJ...834L...8M}, the square is from the VLA \citep{2017Natur.541...58C} and the black stars are detections with MeerKAT. Around MJD 57650, there are observations at L-band with both the VLA and EVN (European VLBI Network) facilities, resulting in two different flux densities, however, we note that the two values still fall within 2$\sigma$ of each other. Furthermore, subsequent VLBI observations obtain similar flux densities as that measured with the VLA indicating that there is no evidence of resolving out of emission by higher angular resolution observations. Therefore we do not expect to measure a different flux density with MeerKAT purely because the array has a different resolving power.

Figure \ref{fig:lightcurve} shows that over a period of nearly seven years, the flux density of the PRS varies: between MJD 57400 and 59850, the average flux density increased from below 180\,$\mu$Jy to around 260\,$\mu$Jy, an increase of around 30\,\%, and then decreased again back to the level measured about MJD 57400. Given that all the flux density measurements presented in this paper are obtained using the same telescope, it motivates the idea that the changes in flux density of the PRS are real. 

Variability was also observed at 3\,GHz by \citet{2017Natur.541...58C}, where the PRS varied between $\sim150-240$\,$\mu$Jy, (shown with a purple shaded region in Figure \ref{fig:lightcurve}). Subsequent work concluded that the likely origin of the variability was scintillation \citep{2017ApJ...842...34W}. We find that the increased flux density observed at 3\,GHz is spectrally consistent with the flux densities observed with MeerKAT i.e. extrapolating the brightest MeerKAT flux densities from 1.3 to 3\,GHz using our measured spectral index results ($\alpha = -0.4\pm0.5$) in measurements that are consistent with the brightest flux densities measured at 3\,GHz. Unfortunately, there are no high-cadence 1.3\,GHz observations during the VLA observing period. 

The variability observed at 3 and 1.3\,GHz probes very different timescales: days/weeks vs years. Using the model presented in \citet{2019arXiv190708395H}, we estimate that the modulation amplitude due to refractive scintillation along the line of sight to the PRS is approximately 25\%, similar to the observed amplitude. The predicted variability timescale is far shorter ($\sim$0.5\,years) than what we observe making it unlikely that scintillation is the cause of the variability. However, we note that there is also the possibility that we were unlucky. Given the low cadence of our observations, there is a chance that we happened to observe when the variability due to scintillation was at a low point. There are also large uncertainties in the models used to calculate the observed effects of scintillation and so our quoted value of the scintillation timescale is only approximate. In order to completely rule out the possibility of scintillation, we need more observations of the PRS over a range of timescales.

Despite the spectral and luminosity similarities, the PWN models \citep{2017ApJ...841...14M, 2019ApJ...885..149Y} developed to explain the PRS do not predict the variability that has been observed at 3 and now 1.3\,GHz \citep{2017Natur.541...58C}. Other PRSs have been discovered too recently to search for variability on the timescale observed in FRB 20121102A. If the variability is intrinsic to the source, and not caused by scintillation, the increase in flux density is particularly hard to explain within the PWN scenario. Given the size constraints \citep[$<0.7$pc][]{2017ApJ...841...14M}, we propose that the varying PRS is from the termination shock. The shock is produced at the interface between the supernova ejecta and the pulsar wind and produces synchrotron radiation. In the case of the Crab Nebula and 3C 58, the termination shock radius is less than a parsec in diameter - consistent with our observations of the PRS \citep{green1992, 2002ApJ...571L..45S, 1984ApJ...283..694K}. If the system that produces FRB 20121102A is particularly young, as indicated by the spectral break at $\sim$8\,GHz \citep{2020arXiv201014334R}, then it is possible that flux variability originates from variation in the termination shock. This could be the result of interaction between the pulsar wind and a reverse shock from the supernova that produced the magnetar in the first place. 

\subsection{FRB detections}

In Section \ref{sub:result_FRB}, we demonstrated that three bursts were bright enough to be detected when imaging the whole band. A further four were detected in one-half of the band. The same dataset was imaged by \citet{2023MNRAS.518.3462A}, who reported detections of bursts two, three, five, seven, eight and eleven. Despite not detecting burst two, we recover an additional two image plane bursts: nine and ten.

\citet{2023MNRAS.518.3462A} provides a clear methodology for detecting faint bursts in MeerKAT images. There are two main differences between the method presented in \citet{2023MNRAS.518.3462A} and what we present here. The first is the software used to perform flagging: here we use \textsc{tricolour} \citep{2022ASPC..532..541H} and keep observatory-placed flags, whereas \citet{2023MNRAS.518.3462A} removes all flags put in place by the observatory and then uses \textsc{aoflagger} and \textsc{casa} \citep{offringa-2012-morph-rfi-algorithm, casa}. \citet{2023MNRAS.518.3462A} explored using \textsc{tricolour} and found that the two methods resulted in different S/Ns for the detected bursts. The second difference is the use of peeling. Peeling is a method by which bright, contaminating sources are removed from the data in the \textit{uv}-plane resulting in a lower rms noise in the final image \citep{2004SPIE.5489..817N,2009A&A...501.1185I}. For example in low frequency subband two-second images, we reach an rms noise of 600$\mu$Jy whereas with peeling \citet{2023MNRAS.518.3462A} reaches 400$\mu$Jy.

In this work, we consider only the bursts that are bright enough to be detected in an image without peeling. We use this methodology to obtain the clearest understanding of the population of bursts that could be detected by commensal searches of MeerKAT fields. Commensal searches to look for new transients and variable sources in a number of ThunderKAT Large Survey Project \citep{2016mks..confE..13F} fields have already been performed \citep{2020MNRAS.491..560D, 2022MNRAS.517.2894R, 2022MNRAS.512.5037D, 2023MNRAS.518.3462A}. These searches were performed by looking for variations in flux density of point sources from epoch to epoch. It is also possible to look for transients and variables, such as pulsars and FRBs, on timescales of seconds by imaging each integration of a whole observation \citep[e.g.][]{2022NatAs...6..828C}. Compared to long-duration transients and variables, it is harder to determine if an FRB i.e. a single bright detection, is real and not an artefact. Through the work performed on other data sets searching for commensal transients, \citet{2023MNRAS.523.2219A} and Chastain et al (\textit{submitted}) have developed a methodology which rules out imaging artefacts such as sidelobes associated with very bright sources, or satellites. As such, with any future FRB candidate, we are confident in our ability to identify it as an FRB and not a result of poor UV coverage or satellite contamination. 

\begin{figure*}
    \centering
    \includegraphics[width = \textwidth]{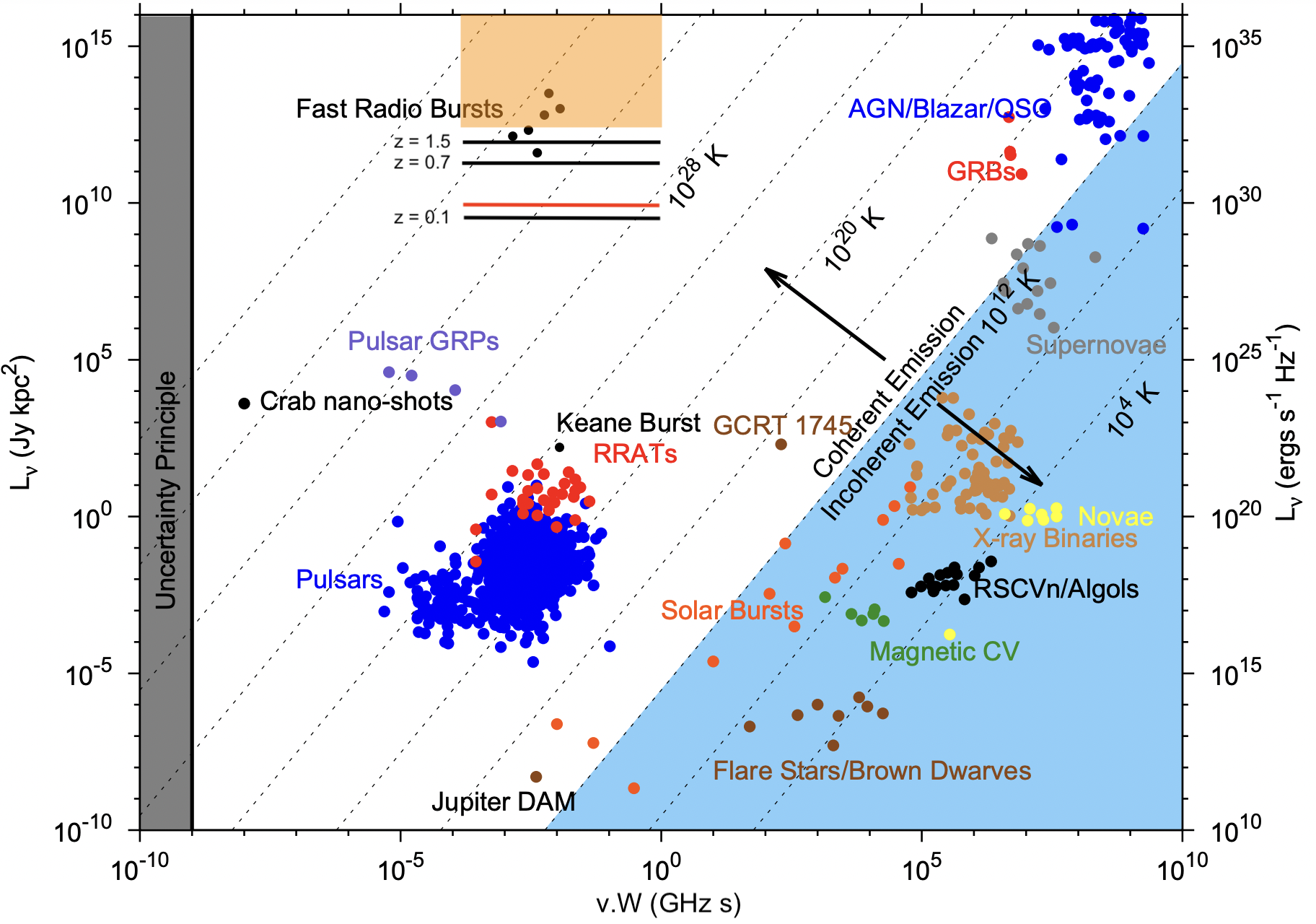}
    \caption{Adapted from \citet{2015MNRAS.446.3687P, 2018NatAs...2..865K}, a plot of the radio luminosity of different radio transients as a function of their variability timescale. The extragalactic FRB population (black data points) spans five orders of magnitude in luminosity space. Also shown is the luminosity distribution for bursts detected from the galactic magnetar SGR 1935$+$2154, which hints at overlapping with the extragalactic FRB population. The three horizontal black lines correspond to the minimum luminosity required for us to detect a burst at z = 0.1, 0.7 and 1.5, in a two-second image. The limit we use for this calculation is an 8$\sigma$ detection which corresponds to a flux density of 2.4\,mJy. The orange-shaded region denotes the luminosity distribution of the CHIME bursts \citep{2022Ap&SS.367...66C}. The red horizontal line indicates the average luminosity of the bursts that we detect in the image plane.}
    \label{fig:pop_frb}
\end{figure*}

The average rms noise in an two-second image is $\sim0.3$\,mJy. Therefore, an FRB would have to have a flux density of at least 2.4\,mJy in a two-second image to be detected in the TraP (the default threshold for a detection to be considered significant in the TraP is 8$\sigma$). Bursts from FRBs usually only last milliseconds and the measured fluence that is usually reported will be smeared out \citep{2013ApJ...767....4T}. Figure \ref{fig:pop_frb} shows the luminosity-timescale parameter space for different transients considering both coherent (pulsars and FRBs) and incoherent transients \citep[e.g. cataclysmic variables and supernovae,][]{2015MNRAS.446.3687P}. Overlaid is a series of horizontal lines which correspond to the luminosities required for an FRB to be detected from a distance of z = 0.1, 0.7 and 1.5. The red line in Figure \ref{fig:pop_frb} is the average luminosity of the bursts reported in this work across the whole bandwidth at the distance of FRB 20121102A. The orange box corresponds to the luminosity distribution of bursts detected by the CHIME telescope \citep{2022Ap&SS.367...66C}. The bursts we detect from FRB 20121102A have very low luminosities compared to the CHIME population. The rms noise values we are able to reach in a two-second observation are low enough that we are confident in our ability to detect, and immediately localise, a large population of FRBs including all bursts with CHIME-like fluences. 

We note that despite being able to localise FRBs with redshifts as high as $z = 1.5$, the uncertainty on the FRB position is dependent on the S/N which means that for bursts at higher redshifts, which may be fainter, the localisation region on the sky could contain multiple possible host galaxies compared to their low-redshift counterparts. The localisation precision of a given source is given as $\approx \mathrm{beamwidth}/\sqrt{S/N}$. For a given burst from FRB 121102, the localisation region is $\sim8"\times2"$ (we note that the beam size is highly elliptical due to the high declination of FRB 121102's position with respect to MeerKAT's observing range), which, at low redshift ($z \lesssim$ 0.1) is sufficient to confidently associate a burst with a host galaxy but that confidence rapidly decreases whilst moving to higher redshifts \citep{2018ApJ...860...73E}.

Looking forward to the SKA (Square Kilometer Array) era, if we are to discover the most complete sample of FRBs possible, shorter integration times are vital in order to find and localise the fainter population of FRBs with higher S/N and therefore better localisation precision. We find that the scientific benefit of shorter integration times is even stronger when considering the data rates involved, when searching for a fainter population of FRBs using backends such as MeerTRAP, higher time resolution is required and therefore to reduce the data rates, only a small fraction of sky can be observed \citep{2018IAUS..337..406S}\footnote{https://skaafrica.atlassian.net/wiki/spaces/ESDKB/pages/\\1591672833/User+Supplied+Equipment+USE\#TUSE}. Comparatively, in the image plane, the same field of view is observed regardless of the integration time. As result, we are able to search for a fainter population of fast transients over a larger fraction of the sky at a given time in the image plane compared to searching in beam-formed data. 

On the other hand, there is the possibility that the effects of large DM values will smear out the FRB signal such that the signal that is only a few milliseconds in length is smeared out over more than one integration. The benefit of the longer integrations times (e.g. four or eight seconds) in the image plane compared to time-series data is that it automatically collapses any smearing in the time domain that would be a result of large DMs such that we do not need to correct for DM unless the burst is dispersed over a period larger than integration time.


\section{Conclusions}

We have presented MeerKAT observations of the PRS associated with FRB 20121102A as well as a number of image-plane detections of the bursts themselves. We find evidence for long-term ($\sim$years) flux density changes in the PRS, which, when combined with the very small source size constraints, may be a result of interaction between a highly magnetised PWN terminal shock with the surrounding supernova ejecta. Continued monitoring of the PRS associated with FRB 20121102A and others is required to determine whether significant changes in flux density on a timescale of years are a global property of PRSs. We also presented the results of two-second imaging, demonstrating MeerKAT's ability to find and localise fast transients in the image plane. We compare the luminosities of our FRB 20121102A detections, the upper limits of detect bursts at a series of redshifts and the current CHIME population and find that we expect to be able to find a large sample size of bursts through commensal searches.

\section*{Acknowledgements}

The authors thank the anonymous referee for their useful comments. LR acknowledges the support given by the Science and Technology
Facilities Council through an STFC post-doctoral research-associate position. LR also acknowledges the useful conversations had with A. Mitchell and S. Spencer. AA acknowledges the support given by the Science
and Technology Facilities Council through a STFC studentship. BWS acknowledge support from the European Research Council (ERC) under the European Union's Horizon 2020 research and innovation programme (grant agreement No 694745). MC acknowledges support of an Australian Research Council Discovery Early Career Research Award (project number DE220100819) funded by the Australian Government and the Australian Research Council Centre of Excellence for All Sky Astrophysics in 3 Dimensions (ASTRO 3D), through project number CE170100013.

This paper uses MeerKAT observations obtained through a Director’s Discretionary Time (DDT) proposal (Project ID: DDT-20190905-MC-01). The MeerKAT telescope is operated by the South African Radio Astronomy Observatory, which is a facility of the National Research Foundation, an agency of the Department of Science and Innovation (DSI). The authors acknowledge the contribution of all those who designed and built the MeerKAT instrument.

We acknowledge the use of the ilifu cloud computing facility - www.ilifu.ac.za, a partnership between the University of Cape Town, the University of the Western Cape, the University of Stellenbosch, Sol Plaatje University, the Cape Peninsula University of Technology and the South African Radio Astronomy Observatory. The ilifu facility is supported by contributions from the Inter-University Institute for Data Intensive Astronomy (IDIA - a partnership between the University of Cape Town, the University of Pretoria and the University of the Western Cape), the Computational Biology division at UCT and the Data Intensive Research Initiative of South Africa (DIRISA).

\section*{Data Availability}

The data presented in this paper will be provided upon request.




\bibliographystyle{mnras}
\bibliography{example} 

\begin{thebibliography}{}
\makeatletter
\relax
\def\mn@urlcharsother{\let\do\@makeother \do\$\do\&\do\#\do\^\do\_\do\%\do\~}
\def\mn@doi{\begingroup\mn@urlcharsother \@ifnextchar [ {\mn@doi@}
  {\mn@doi@[]}}
\def\mn@doi@[#1]#2{\def\@tempa{#1}\ifx\@tempa\@empty \href
  {http://dx.doi.org/#2} {doi:#2}\else \href {http://dx.doi.org/#2} {#1}\fi
  \endgroup}
\def\mn@eprint#1#2{\mn@eprint@#1:#2::\@nil}
\def\mn@eprint@arXiv#1{\href {http://arxiv.org/abs/#1} {{\tt arXiv:#1}}}
\def\mn@eprint@dblp#1{\href {http://dblp.uni-trier.de/rec/bibtex/#1.xml}
  {dblp:#1}}
\def\mn@eprint@#1:#2:#3:#4\@nil{\def\@tempa {#1}\def\@tempb {#2}\def\@tempc
  {#3}\ifx \@tempc \@empty \let \@tempc \@tempb \let \@tempb \@tempa \fi \ifx
  \@tempb \@empty \def\@tempb {arXiv}\fi \@ifundefined
  {mn@eprint@\@tempb}{\@tempb:\@tempc}{\expandafter \expandafter \csname
  mn@eprint@\@tempb\endcsname \expandafter{\@tempc}}}

\bibitem[\protect\citeauthoryear{{Andersson} et~al.,}{{Andersson}
  et~al.}{2022}]{2022MNRAS.513.3482A}
{Andersson} A.,  et~al., 2022, \mn@doi [\mnras] {10.1093/mnras/stac1002}, \href
  {https://ui.adsabs.harvard.edu/abs/2022MNRAS.513.3482A} {513, 3482}

\bibitem[\protect\citeauthoryear{{Andersson} et~al.,}{{Andersson}
  et~al.}{2023}]{2023MNRAS.523.2219A}
{Andersson} A.,  et~al., 2023, \mn@doi [\mnras] {10.1093/mnras/stad1298}, \href
  {https://ui.adsabs.harvard.edu/abs/2023MNRAS.523.2219A} {523, 2219}

\bibitem[\protect\citeauthoryear{{Andrianjafy} et~al.,}{{Andrianjafy}
  et~al.}{2023}]{2023MNRAS.518.3462A}
{Andrianjafy} J.~C.,  et~al., 2023, \mn@doi [\mnras] {10.1093/mnras/stac3348},
  \href {https://ui.adsabs.harvard.edu/abs/2023MNRAS.518.3462A} {518, 3462}

\bibitem[\protect\citeauthoryear{{Bochenek}, {Ravi}, {Belov}, {Hallinan},
  {Kocz}, {Kulkarni}  \& {McKenna}}{{Bochenek}
  et~al.}{2020}]{2020Natur.587...59B}
{Bochenek} C.~D.,  {Ravi} V.,  {Belov} K.~V.,  {Hallinan} G.,  {Kocz} J.,
  {Kulkarni} S.~R.,   {McKenna} D.~L.,  2020, \mn@doi [\nat]
  {10.1038/s41586-020-2872-x}, \href
  {https://ui.adsabs.harvard.edu/abs/2020Natur.587...59B} {587, 59}

\bibitem[\protect\citeauthoryear{{Bucciantini}, {Arons}  \&
  {Amato}}{{Bucciantini} et~al.}{2011}]{2011MNRAS.410..381B}
{Bucciantini} N.,  {Arons} J.,   {Amato} E.,  2011, \mn@doi [\mnras]
  {10.1111/j.1365-2966.2010.17449.x}, \href
  {https://ui.adsabs.harvard.edu/abs/2011MNRAS.410..381B} {410, 381}

\bibitem[\protect\citeauthoryear{{CHIME/FRB Collaboration} et~al.,}{{CHIME/FRB
  Collaboration} et~al.}{2019}]{2019ApJ...885L..24C}
{CHIME/FRB Collaboration} et~al., 2019, \mn@doi [\apjl]
  {10.3847/2041-8213/ab4a80}, \href
  {https://ui.adsabs.harvard.edu/abs/2019ApJ...885L..24C} {885, L24}

\bibitem[\protect\citeauthoryear{{Caleb} et~al.,}{{Caleb}
  et~al.}{2020}]{2020MNRAS.496.4565C}
{Caleb} M.,  et~al., 2020, \mn@doi [\mnras] {10.1093/mnras/staa1791}, \href
  {https://ui.adsabs.harvard.edu/abs/2020MNRAS.496.4565C} {496, 4565}

\bibitem[\protect\citeauthoryear{{Caleb} et~al.,}{{Caleb}
  et~al.}{2022}]{2022NatAs...6..828C}
{Caleb} M.,  et~al., 2022, \mn@doi [Nature Astronomy]
  {10.1038/s41550-022-01688-x}, \href
  {https://ui.adsabs.harvard.edu/abs/2022NatAs...6..828C} {6, 828}

\bibitem[\protect\citeauthoryear{{Chatterjee} et~al.,}{{Chatterjee}
  et~al.}{2017}]{2017Natur.541...58C}
{Chatterjee} S.,  et~al., 2017, \mn@doi [\nat] {10.1038/nature20797}, \href
  {https://ui.adsabs.harvard.edu/abs/2017Natur.541...58C} {541, 58}

\bibitem[\protect\citeauthoryear{{Chibueze} et~al.,}{{Chibueze}
  et~al.}{2022}]{2022MNRAS.515.1365C}
{Chibueze} J.~O.,  et~al., 2022, \mn@doi [\mnras] {10.1093/mnras/stac1601},
  \href {https://ui.adsabs.harvard.edu/abs/2022MNRAS.515.1365C} {515, 1365}

\bibitem[\protect\citeauthoryear{{Chime/Frb Collaboration} et~al.,}{{Chime/Frb
  Collaboration} et~al.}{2020}]{2020Natur.582..351C}
{Chime/Frb Collaboration} et~al., 2020, \mn@doi [\nat]
  {10.1038/s41586-020-2398-2}, \href
  {https://ui.adsabs.harvard.edu/abs/2020Natur.582..351C} {582, 351}

\bibitem[\protect\citeauthoryear{{Cruces} et~al.,}{{Cruces}
  et~al.}{2021}]{2021MNRAS.500..448C}
{Cruces} M.,  et~al., 2021, \mn@doi [\mnras] {10.1093/mnras/staa3223}, \href
  {https://ui.adsabs.harvard.edu/abs/2021MNRAS.500..448C} {500, 448}

\bibitem[\protect\citeauthoryear{{Cui} et~al.,}{{Cui}
  et~al.}{2022}]{2022Ap&SS.367...66C}
{Cui} X.-H.,  et~al., 2022, \mn@doi [\apss] {10.1007/s10509-022-04093-y}, \href
  {https://ui.adsabs.harvard.edu/abs/2022Ap&SS.367...66C} {367, 66}

\bibitem[\protect\citeauthoryear{{Driessen} et~al.,}{{Driessen}
  et~al.}{2020}]{2020MNRAS.491..560D}
{Driessen} L.~N.,  et~al., 2020, \mn@doi [\mnras] {10.1093/mnras/stz3027},
  \href {https://ui.adsabs.harvard.edu/abs/2020MNRAS.491..560D} {491, 560}

\bibitem[\protect\citeauthoryear{{Driessen} et~al.,}{{Driessen}
  et~al.}{2022}]{2022MNRAS.512.5037D}
{Driessen} L.~N.,  et~al., 2022, \mn@doi [\mnras] {10.1093/mnras/stac756},
  \href {https://ui.adsabs.harvard.edu/abs/2022MNRAS.512.5037D} {512, 5037}

\bibitem[\protect\citeauthoryear{{Driessen} et~al.,}{{Driessen}
  et~al.}{2023}]{2023arXiv230209787D}
{Driessen} L.~N.,  et~al., 2023, \mn@doi [arXiv e-prints]
  {10.48550/arXiv.2302.09787}, \href
  {https://ui.adsabs.harvard.edu/abs/2023arXiv230209787D} {p. arXiv:2302.09787}

\bibitem[\protect\citeauthoryear{{Dubner}, {Giacani}  \&
  {Decourchelle}}{{Dubner} et~al.}{2008}]{2008A&A...487.1033D}
{Dubner} G.,  {Giacani} E.,   {Decourchelle} A.,  2008, \mn@doi [\aap]
  {10.1051/0004-6361:200809987}, \href
  {https://ui.adsabs.harvard.edu/abs/2008A&A...487.1033D} {487, 1033}

\bibitem[\protect\citeauthoryear{{Eftekhari}, {Berger}, {Williams}  \&
  {Blanchard}}{{Eftekhari} et~al.}{2018}]{2018ApJ...860...73E}
{Eftekhari} T.,  {Berger} E.,  {Williams} P.~K.~G.,   {Blanchard} P.~K.,  2018,
  \mn@doi [\apj] {10.3847/1538-4357/aac270}, \href
  {https://ui.adsabs.harvard.edu/abs/2018ApJ...860...73E} {860, 73}

\bibitem[\protect\citeauthoryear{{Fender} et~al.,}{{Fender}
  et~al.}{2016}]{2016mks..confE..13F}
{Fender} R.,  et~al., 2016, in MeerKAT Science: On the Pathway to the SKA.
  p.~13 (\mn@eprint {arXiv} {1711.04132}), \mn@doi{10.22323/1.277.0013}

\bibitem[\protect\citeauthoryear{{Fonseca} et~al.,}{{Fonseca}
  et~al.}{2020}]{2020ApJ...891L...6F}
{Fonseca} E.,  et~al., 2020, \mn@doi [\apjl] {10.3847/2041-8213/ab7208}, \href
  {https://ui.adsabs.harvard.edu/abs/2020ApJ...891L...6F} {891, L6}

\bibitem[\protect\citeauthoryear{{Gaensler} \& {Slane}}{{Gaensler} \&
  {Slane}}{2006}]{2006ARA&A..44...17G}
{Gaensler} B.~M.,  {Slane} P.~O.,  2006, \mn@doi [\araa]
  {10.1146/annurev.astro.44.051905.092528}, \href
  {https://ui.adsabs.harvard.edu/abs/2006ARA&A..44...17G} {44, 17}

\bibitem[\protect\citeauthoryear{{Green} \& {Scheuer}}{{Green} \&
  {Scheuer}}{1992}]{green1992}
{Green} D.~A.,  {Scheuer} P.~A.~G.,  1992, \mn@doi [\mnras]
  {10.1093/mnras/258.4.833}, \href
  {https://ui.adsabs.harvard.edu/abs/1992MNRAS.258..833G} {258, 833}

\bibitem[\protect\citeauthoryear{{Hancock}, {Charlton}, {Macquart}  \&
  {Hurley-Walker}}{{Hancock} et~al.}{2019}]{2019arXiv190708395H}
{Hancock} P.~J.,  {Charlton} E.~G.,  {Macquart} J.-P.,   {Hurley-Walker} N.,
  2019, arXiv e-prints, \href
  {https://ui.adsabs.harvard.edu/abs/2019arXiv190708395H} {p. arXiv:1907.08395}

\bibitem[\protect\citeauthoryear{{Hugo}, {Perkins}, {Merry}, {Mauch}  \&
  {Smirnov}}{{Hugo} et~al.}{2022}]{2022ASPC..532..541H}
{Hugo} B.~V.,  {Perkins} S.,  {Merry} B.,  {Mauch} T.,   {Smirnov} O.~M.,
  2022, in {Ruiz} J.~E.,  {Pierfedereci} F.,   {Teuben} P.,  eds,  Astronomical
  Society of the Pacific Conference Series Vol. 532, Astronomical Society of
  the Pacific Conference Series. p.~541 (\mn@eprint {arXiv} {2206.09179}),
  \mn@doi{10.48550/arXiv.2206.09179}

\bibitem[\protect\citeauthoryear{{Intema}, {van der Tol}, {Cotton}, {Cohen},
  {van Bemmel}  \& {R{\"o}ttgering}}{{Intema}
  et~al.}{2009}]{2009A&A...501.1185I}
{Intema} H.~T.,  {van der Tol} S.,  {Cotton} W.~D.,  {Cohen} A.~S.,  {van
  Bemmel} I.~M.,   {R{\"o}ttgering} H.~J.~A.,  2009, \mn@doi [\aap]
  {10.1051/0004-6361/200811094}, \href
  {https://ui.adsabs.harvard.edu/abs/2009A&A...501.1185I} {501, 1185}

\bibitem[\protect\citeauthoryear{{Jonas} \& {MeerKAT Team}}{{Jonas} \& {MeerKAT
  Team}}{2016}]{2016mks..confE...1J}
{Jonas} J.,  {MeerKAT Team} 2016, in MeerKAT Science: On the Pathway to the
  SKA. p.~1, \mn@doi{10.22323/1.277.0001}

\bibitem[\protect\citeauthoryear{{Keane}}{{Keane}}{2018}]{2018NatAs...2..865K}
{Keane} E.~F.,  2018, \mn@doi [Nature Astronomy] {10.1038/s41550-018-0603-0},
  \href {https://ui.adsabs.harvard.edu/abs/2018NatAs...2..865K} {2, 865}

\bibitem[\protect\citeauthoryear{{Keane}, {Stappers}, {Kramer}  \&
  {Lyne}}{{Keane} et~al.}{2012}]{2012MNRAS.425L..71K}
{Keane} E.~F.,  {Stappers} B.~W.,  {Kramer} M.,   {Lyne} A.~G.,  2012, \mn@doi
  [\mnras] {10.1111/j.1745-3933.2012.01306.x}, \href
  {https://ui.adsabs.harvard.edu/abs/2012MNRAS.425L..71K} {425, L71}

\bibitem[\protect\citeauthoryear{{Kennel} \& {Coroniti}}{{Kennel} \&
  {Coroniti}}{1984}]{1984ApJ...283..694K}
{Kennel} C.~F.,  {Coroniti} F.~V.,  1984, \mn@doi [\apj] {10.1086/162356},
  \href {https://ui.adsabs.harvard.edu/abs/1984ApJ...283..694K} {283, 694}

\bibitem[\protect\citeauthoryear{{Law}, {Connor}  \& {Aggarwal}}{{Law}
  et~al.}{2022}]{2022ApJ...927...55L}
{Law} C.~J.,  {Connor} L.,   {Aggarwal} K.,  2022, \mn@doi [\apj]
  {10.3847/1538-4357/ac4c42}, \href
  {https://ui.adsabs.harvard.edu/abs/2022ApJ...927...55L} {927, 55}

\bibitem[\protect\citeauthoryear{{Levin}, {Beloborodov}  \&
  {Bransgrove}}{{Levin} et~al.}{2020}]{2020ApJ...895L..30L}
{Levin} Y.,  {Beloborodov} A.~M.,   {Bransgrove} A.,  2020, \mn@doi [\apjl]
  {10.3847/2041-8213/ab8c4c}, \href
  {https://ui.adsabs.harvard.edu/abs/2020ApJ...895L..30L} {895, L30}

\bibitem[\protect\citeauthoryear{{Lorimer}, {Bailes}, {McLaughlin}, {Narkevic}
  \& {Crawford}}{{Lorimer} et~al.}{2007}]{2007Sci...318..777L}
{Lorimer} D.~R.,  {Bailes} M.,  {McLaughlin} M.~A.,  {Narkevic} D.~J.,
  {Crawford} F.,  2007, \mn@doi [Science] {10.1126/science.1147532}, \href
  {https://ui.adsabs.harvard.edu/abs/2007Sci...318..777L} {318, 777}

\bibitem[\protect\citeauthoryear{{Lyutikov}, {Barkov}  \&
  {Giannios}}{{Lyutikov} et~al.}{2020}]{2020ApJ...893L..39L}
{Lyutikov} M.,  {Barkov} M.~V.,   {Giannios} D.,  2020, \mn@doi [\apjl]
  {10.3847/2041-8213/ab87a4}, \href
  {https://ui.adsabs.harvard.edu/abs/2020ApJ...893L..39L} {893, L39}

\bibitem[\protect\citeauthoryear{{Maitra}, {Roy}, {Acero}  \& {Gupta}}{{Maitra}
  et~al.}{2018}]{2018MNRAS.477L..66M}
{Maitra} C.,  {Roy} S.,  {Acero} F.,   {Gupta} Y.,  2018, \mn@doi [\mnras]
  {10.1093/mnrasl/sly038}, \href
  {https://ui.adsabs.harvard.edu/abs/2018MNRAS.477L..66M} {477, L66}

\bibitem[\protect\citeauthoryear{{Marcote} et~al.,}{{Marcote}
  et~al.}{2017}]{2017ApJ...834L...8M}
{Marcote} B.,  et~al., 2017, \mn@doi [\apjl] {10.3847/2041-8213/834/2/L8},
  \href {https://ui.adsabs.harvard.edu/abs/2017ApJ...834L...8M} {834, L8}

\bibitem[\protect\citeauthoryear{{Marcote} et~al.,}{{Marcote}
  et~al.}{2021}]{2021ATel14603....1M}
{Marcote} B.,  et~al., 2021, The Astronomer's Telegram, \href
  {https://ui.adsabs.harvard.edu/abs/2021ATel14603....1M} {14603, 1}

\bibitem[\protect\citeauthoryear{{McMullin}, {Waters}, {Schiebel}, {Young}  \&
  {Golap}}{{McMullin} et~al.}{2007}]{casa}
{McMullin} J.~P.,  {Waters} B.,  {Schiebel} D.,  {Young} W.,   {Golap} K.,
  2007, {CASA Architecture and Applications}.
Astron. Soc. Pac., San Francisco, p.~127

\bibitem[\protect\citeauthoryear{{Metzger}, {Berger}  \& {Margalit}}{{Metzger}
  et~al.}{2017}]{2017ApJ...841...14M}
{Metzger} B.~D.,  {Berger} E.,   {Margalit} B.,  2017, \mn@doi [\apj]
  {10.3847/1538-4357/aa633d}, \href
  {https://ui.adsabs.harvard.edu/abs/2017ApJ...841...14M} {841, 14}

\bibitem[\protect\citeauthoryear{{Niu} et~al.,}{{Niu}
  et~al.}{2022}]{2022Natur.606..873N}
{Niu} C.~H.,  et~al., 2022, \mn@doi [\nat] {10.1038/s41586-022-04755-5}, \href
  {https://ui.adsabs.harvard.edu/abs/2022Natur.606..873N} {606, 873}

\bibitem[\protect\citeauthoryear{{Noordam}}{{Noordam}}{2004}]{2004SPIE.5489..817N}
{Noordam} J.~E.,  2004, in {Oschmann} Jacobus~M. J.,  ed.,  Society of
  Photo-Optical Instrumentation Engineers (SPIE) Conference Series Vol. 5489,
  Ground-based Telescopes. pp 817--825, \mn@doi{10.1117/12.544262}

\bibitem[\protect\citeauthoryear{Offringa, van~de Gronde  \& Roerdink}{Offringa
  et~al.}{2012}]{offringa-2012-morph-rfi-algorithm}
Offringa A.~R.,  van~de Gronde J.~J.,   Roerdink J. B. T.~M.,  2012, A\&A, 539

\bibitem[\protect\citeauthoryear{{Offringa} et~al.,}{{Offringa}
  et~al.}{2014}]{2014MNRAS.444..606O}
{Offringa} A.~R.,  et~al., 2014, \mn@doi [\mnras] {10.1093/mnras/stu1368},
  \href {https://ui.adsabs.harvard.edu/abs/2014MNRAS.444..606O} {444, 606}

\bibitem[\protect\citeauthoryear{{Petroff} et~al.,}{{Petroff}
  et~al.}{2016}]{2016PASA...33...45P}
{Petroff} E.,  et~al., 2016, \mn@doi [\pasa] {10.1017/pasa.2016.35}, \href
  {https://ui.adsabs.harvard.edu/abs/2016PASA...33...45P} {33, e045}

\bibitem[\protect\citeauthoryear{{Pietka}, {Fender}  \& {Keane}}{{Pietka}
  et~al.}{2015}]{2015MNRAS.446.3687P}
{Pietka} M.,  {Fender} R.~P.,   {Keane} E.~F.,  2015, \mn@doi [\mnras]
  {10.1093/mnras/stu2335}, \href
  {https://ui.adsabs.harvard.edu/abs/2015MNRAS.446.3687P} {446, 3687}

\bibitem[\protect\citeauthoryear{{Platts} et~al.,}{{Platts}
  et~al.}{2021}]{2021MNRAS.505.3041P}
{Platts} E.,  et~al., 2021, \mn@doi [\mnras] {10.1093/mnras/stab1544}, \href
  {https://ui.adsabs.harvard.edu/abs/2021MNRAS.505.3041P} {505, 3041}

\bibitem[\protect\citeauthoryear{{Plavin}, {Paragi}, {Marcote}, {Keimpema},
  {Hessels}, {Nimmo}, {Vedantham}  \& {Spitler}}{{Plavin}
  et~al.}{2022}]{2022MNRAS.511.6033P}
{Plavin} A.,  {Paragi} Z.,  {Marcote} B.,  {Keimpema} A.,  {Hessels} J.~W.~T.,
  {Nimmo} K.,  {Vedantham} H.~K.,   {Spitler} L.~G.,  2022, \mn@doi [\mnras]
  {10.1093/mnras/stac500}, \href
  {https://ui.adsabs.harvard.edu/abs/2022MNRAS.511.6033P} {511, 6033}

\bibitem[\protect\citeauthoryear{{Rajwade} et~al.,}{{Rajwade}
  et~al.}{2020}]{2020MNRAS.495.3551R}
{Rajwade} K.~M.,  et~al., 2020, \mn@doi [\mnras] {10.1093/mnras/staa1237},
  \href {https://ui.adsabs.harvard.edu/abs/2020MNRAS.495.3551R} {495, 3551}

\bibitem[\protect\citeauthoryear{{Ravi} et~al.,}{{Ravi}
  et~al.}{2022}]{2022MNRAS.513..982R}
{Ravi} V.,  et~al., 2022, \mn@doi [\mnras] {10.1093/mnras/stac465}, \href
  {https://ui.adsabs.harvard.edu/abs/2022MNRAS.513..982R} {513, 982}

\bibitem[\protect\citeauthoryear{{Resmi}, {Vink}  \& {Ishwara-Chandra}}{{Resmi}
  et~al.}{2021}]{2020arXiv201014334R}
{Resmi} L.,  {Vink} J.,   {Ishwara-Chandra} C.~H.,  2021, \mn@doi [\aap]
  {10.1051/0004-6361/202039771}, \href
  {https://ui.adsabs.harvard.edu/abs/2021A&A...655A.102R} {655, A102}

\bibitem[\protect\citeauthoryear{{Reynolds} \& {Chevalier}}{{Reynolds} \&
  {Chevalier}}{1984}]{1984ApJ...278..630R}
{Reynolds} S.~P.,  {Chevalier} R.~A.,  1984, \mn@doi [\apj] {10.1086/161831},
  \href {https://ui.adsabs.harvard.edu/abs/1984ApJ...278..630R} {278, 630}

\bibitem[\protect\citeauthoryear{{Rowlinson} et~al.,}{{Rowlinson}
  et~al.}{2022}]{2022MNRAS.517.2894R}
{Rowlinson} A.,  et~al., 2022, \mn@doi [\mnras] {10.1093/mnras/stac2460}, \href
  {https://ui.adsabs.harvard.edu/abs/2022MNRAS.517.2894R} {517, 2894}

\bibitem[\protect\citeauthoryear{Sanidas, Caleb, Driessen, Morello, Rajwade  \&
  Stappers}{Sanidas et~al.}{2017}]{2018IAUS..337..406S}
Sanidas S.,  Caleb M.,  Driessen L.,  Morello V.,  Rajwade K.,   Stappers
  B.~W.,  2017, \mn@doi [Proceedings of the International Astronomical Union]
  {10.1017/S1743921317009310}, 13, 406–407

\bibitem[\protect\citeauthoryear{{Scholz} et~al.,}{{Scholz}
  et~al.}{2016}]{2016ApJ...833..177S}
{Scholz} P.,  et~al., 2016, \mn@doi [\apj] {10.3847/1538-4357/833/2/177}, \href
  {https://ui.adsabs.harvard.edu/abs/2016ApJ...833..177S} {833, 177}

\bibitem[\protect\citeauthoryear{{Slane}, {Helfand}  \& {Murray}}{{Slane}
  et~al.}{2002}]{2002ApJ...571L..45S}
{Slane} P.~O.,  {Helfand} D.~J.,   {Murray} S.~S.,  2002, \mn@doi [\apjl]
  {10.1086/341179}, \href
  {https://ui.adsabs.harvard.edu/abs/2002ApJ...571L..45S} {571, L45}

\bibitem[\protect\citeauthoryear{{Spitler} et~al.,}{{Spitler}
  et~al.}{2016}]{2016Natur.531..202S}
{Spitler} L.~G.,  et~al., 2016, \mn@doi [\nat] {10.1038/nature17168}, \href
  {https://ui.adsabs.harvard.edu/abs/2016Natur.531..202S} {531, 202}

\bibitem[\protect\citeauthoryear{{Stappers}}{{Stappers}}{2016}]{2016mks..confE..10S}
{Stappers} B.,  2016, in MeerKAT Science: On the Pathway to the SKA. p.~10,
  \mn@doi{10.22323/1.277.0010}

\bibitem[\protect\citeauthoryear{{Swinbank} et~al.,}{{Swinbank}
  et~al.}{2015}]{2015A&C....11...25S}
{Swinbank} J.~D.,  et~al., 2015, \mn@doi [Astronomy and Computing]
  {10.1016/j.ascom.2015.03.002}, \href
  {https://ui.adsabs.harvard.edu/abs/2015A&C....11...25S} {11, 25}

\bibitem[\protect\citeauthoryear{{Tendulkar} et~al.,}{{Tendulkar}
  et~al.}{2017}]{2017ApJ...834L...7T}
{Tendulkar} S.~P.,  et~al., 2017, \mn@doi [\apjl] {10.3847/2041-8213/834/2/L7},
  \href {https://ui.adsabs.harvard.edu/abs/2017ApJ...834L...7T} {834, L7}

\bibitem[\protect\citeauthoryear{{The Chime/Frb Collaboration} Andersen
  et~al.,}{{The Chime/Frb Collaboration} et~al.}{2020}]{2020Natur.587...54T}
{The Chime/Frb Collaboration} Andersen B. {\^A}.~C.,  et~al., 2020, \mn@doi
  [\nat] {10.1038/s41586-020-2863-y}, \href
  {https://ui.adsabs.harvard.edu/abs/2020Natur.587...54T} {587, 54}

\bibitem[\protect\citeauthoryear{{Thornton} et~al.,}{{Thornton}
  et~al.}{2013}]{2013Sci...341...53T}
{Thornton} D.,  et~al., 2013, \mn@doi [Science] {10.1126/science.1236789},
  \href {https://ui.adsabs.harvard.edu/abs/2013Sci...341...53T} {341, 53}

\bibitem[\protect\citeauthoryear{{Trott} et~al.,}{{Trott}
  et~al.}{2013}]{2013ApJ...767....4T}
{Trott} C.~M.,  et~al., 2013, \mn@doi [\apj] {10.1088/0004-637X/767/1/4}, \href
  {https://ui.adsabs.harvard.edu/abs/2013ApJ...767....4T} {767, 4}

\bibitem[\protect\citeauthoryear{{Waxman}}{{Waxman}}{2017}]{2017ApJ...842...34W}
{Waxman} E.,  2017, \mn@doi [\apj] {10.3847/1538-4357/aa713e}, \href
  {https://ui.adsabs.harvard.edu/abs/2017ApJ...842...34W} {842, 34}

\bibitem[\protect\citeauthoryear{{Yang} \& {Dai}}{{Yang} \&
  {Dai}}{2019}]{2019ApJ...885..149Y}
{Yang} Y.-H.,  {Dai} Z.-G.,  2019, \mn@doi [\apj] {10.3847/1538-4357/ab48dd},
  \href {https://ui.adsabs.harvard.edu/abs/2019ApJ...885..149Y} {885, 149}

\bibitem[\protect\citeauthoryear{{Zanazzi} \& {Lai}}{{Zanazzi} \&
  {Lai}}{2020}]{2020ApJ...892L..15Z}
{Zanazzi} J.~J.,  {Lai} D.,  2020, \mn@doi [\apjl] {10.3847/2041-8213/ab7cdd},
  \href {https://ui.adsabs.harvard.edu/abs/2020ApJ...892L..15Z} {892, L15}

\bibitem[\protect\citeauthoryear{{Zhang}}{{Zhang}}{2020}]{2020Natur.587...45Z}
{Zhang} B.,  2020, \mn@doi [\nat] {10.1038/s41586-020-2828-1}, \href
  {https://ui.adsabs.harvard.edu/abs/2020Natur.587...45Z} {587, 45}

\bibitem[\protect\citeauthoryear{{Zhang}}{{Zhang}}{2022}]{2022arXiv221203972Z}
{Zhang} B.,  2022, \mn@doi [arXiv e-prints] {10.48550/arXiv.2212.03972}, \href
  {https://ui.adsabs.harvard.edu/abs/2022arXiv221203972Z} {p. arXiv:2212.03972}

\makeatother
\end{thebibliography}








\bsp	
\label{lastpage}
\end{document}